# Strong suppression of the Curie temperature in the electron-doped system $La_{1-x}Ce_xCoO_3$

D. Fuchs[1], P. Schweiss[1], P. Adelmann[1], T. Schwarz[1,2], and R. Schneider[1]

[1]Forschungszentrum Karlsruhe, Institut für Festkörperphysik, P. O. Box 3640, D-76021 Karlsruhe, Germany

[2]Universität Karlsruhe, Fakultät für Physik, D-76128 Karlsruhe, Germany

We show for the system $La_{1-x}Ce_xCoO_3$ ($0.1 \leq x \leq 0.4$) that it is possible to synthesize *electron*-doped cobaltites by the growth of epitaxial thin films. For $La_{1-x}Ce_xCoO_3$, ferromagnetic order is observed within the entire doping range (with the maximum of the Curie temperature, $T_c$, at $x \approx 0.3$), resulting in a magnetic phase diagram similar to that of hole-doped lanthanum cobaltites. The measured spin values strongly suggest an intermediate-spin state of the Co ions which has been also found in the hole-doped system. In contrast to the hole-doped material, however, where $T_c$ is well above 200 K, we observe a strong suppression of the maximum $T_c$ to about 22 K. This is likely to be caused by a considerable decrease of the Co3d - O2p hybridization. The observed intriguing magnetic properties are in agreement with previously reported theoretical results.

Hole-doped rare-earth cobaltites of the form $R_{1-x}A_xCoO_3$ (R = trivalent rare-earth element, A = divalent alkaline-earth element) have attracted a lot of attention for quite some time due to their interesting magnetic properties. The evolution of magnetism with the hole-doping level x and with temperature is governed by the transition from the spin-glass-like state to a cluster-glass state or to the ferromagnetic (FM) metallic state[1,2], while double exchange, involving Co ions in different charge states, provides the magnetic coupling. In addition to the lattice, charge, and spin degrees of freedom found in the manganites and many other transition metal oxides, the cobaltites also display a degree of freedom in the spin state of the Co ion. Since the Hund's coupling constant is comparable to the crystal field splitting various spin states are possible, such as low-spin (LS) $t^6_{2g}e^0_g$, S = 0, intermediate-spin (IS) $t^5_{2g}e^1_g$, S = 1, and high-spin state (HS) $t^4_{2g}e^2_g$, S = 2, which are favored by large crystal-field splitting, covalency and exchange energy, respectively[3,4]. The question we want to address in this paper is how electron doping affects the magnetic properties of the cobaltite perovskites.

To this end, we substituted the trivalent rare earth $La^{3+}$ by tetravalent $Ce^{4+}$ ions. To our knowledge, the preparation of single phase bulk material $La_{1-x}Ce_xCoO_3$ by standard solid state reaction has not been achieved up to now. Mitra et al.[5] already succeeded in the preparation of cerium doped manganites by the growth of thin films and demonstrated the electron doping of $La_{0.7}Ce_{0.3}MnO_3$ by x-ray absorption measurements[6]. $La_{0.7}Ce_{0.3}MnO_3$ displays properties remarkably similar to the hole-doped system $La_{0.7}Ca_{0.3}MnO_3$, i. e., a ferromagnetic metallic ground state with a $T_c$ of and a metal-insulator transition at about 250 K[7]. This indicates a similar behavior of the mixed valent states of $Mn^{3+}$-$Mn^{4+}$ (hole doping) and $Mn^{3+}$-$Mn^{2+}$ (electron doping). An intuitive explanation for this is that, due to the splitting of the spin-$e_g$ states of the $Mn^{3+}$ ion ($t^3_ge^1_g$), hole and electron doping x create the same amount of free charge carriers in the $e_g$ band which mediate the double exchange interaction.

In the cobaltites, the situation is more complicated due to the different possible Co spin states. Zhang et al.[8] have carried out theoretical investigations on the magnetic structure of $La_{1-}$ $_x$Ce$_x$CoO$_3$ and postulate a LS state for x < 0.08 and a LS-IS ferromagnetically ordered state for 0.08 ≤ x < 0.83.

In this work, we report on the first synthesis of electron-doped cobaltites by pulsed laser deposition of epitaxial thin films and their intriguing magnetic properties illuminating the spin state in comparison to the corresponding hole-doped cobaltites.

For the pulsed laser deposition of La$_{1-x}$Ce$_x$CoO$_3$ (x = 0.1, 0.2, 0.3 and 0.4) we have used sintered targets which were prepared by the sol-gel-method [9] in order to improve the Ce diffusion and homogeneity of the starting material. However, we did not succeed in the preparation of single phase target material, not even for x = 0.1. For each target we observed an impurity phase of CeO$_2$, increasing with the Ce concentration x. The films were deposited with a thickness of about 80 nm on <001> oriented LaAlO$_3$ single crystal substrates. The growth conditions, i. e., substrate temperature, T$_s$, and laser energy, E, were optimized with respect to crystallinity and magnetic properties of the resulting films.

Despite the inhomogeneity of the targets we succeeded in the growth of highly epitaxial single-phase thin films up to x = 0.4 at T$_s$ = 550 °C, E = 450 mJ, and at an oxygen partial pressure of p(O$_2$) = 0.12 mbar. Fig. 1 demonstrates the epitaxial quality of the <001> oriented films. Only 00$l$ reflections can be observed. The rocking curve at the 002 reflection, see inset of Fig. 1, reveals a mosaic spread smaller than 0.2°. The c-axis lattice parameter, c ≈ 3.84 Å, was found to be nearly constant within the entire doping range and significantly larger than the bulk value, c$_b$ = 3.80 Å, which we determined from the target material. The difference is likely to be caused by an in-plane compressive strain due to the epitaxial growth on LaAlO$_3$ substrates with an in-plane lattice parameter of 3.78 Å. We also carried out transmission electron microscopy and electron diffraction x-ray analysis confirming the microstructural and chemical homogeneity of the samples.

The magnetic properties of the films were studied using a Quantum Design MPMS SQUID system. The zero-field-cooled (ZFC) and field-cooled (FC) magnetization was measured with an external field strength of H = 200 Oe applied parallel to the film surface in the temperature range of 3 K ≤ T ≤ 300 K. In Fig. 2, we display the FC magnetization normalized to the film thickness for x = 0.1, 0.2, 0.3, and 0.4. A transition to a ferromagnetic state is observed for all samples. The Curie temperature, $T_c$, was determined from the onset of the FC magnetization and is shown as a function of the Ce-doping level in the inset of Fig. 2. With increasing electron-doping level x, $T_c$ increases almost linearly up to x = 0.3, where the maximum value of $T_c$ = 22 K is reached. In comparison to the hole-doped cobaltites, where the maximum $T_c$ is 240 K for Sr doping [10], the Curie temperature of the electron-doped samples is strongly suppressed.[11] Nevertheless, it seems to saturate at the same doping concentration x ≈ 0.3, leading to a magnetic phase diagram which is qualitatively similar to that of the hole-doped cobaltites.

However, resistivity measurements indicate that the films are rather in a ferromagnetic insulating than a metallic state. The room temperature resistivity increases with increasing doping level from $\rho \approx 78$ $\Omega$cm for x = 0.1 to $1.8 \times 10^3$ $\Omega$cm for x = 0.3 and $\rho > 4 \times 10^4$ $\Omega$cm for x = 0.4. The increase of localization of charge carriers can be explained in terms of the tolerance factor, t = (A-O)/√2(B-O), where A, B, and O are the ionic radii within the $ABO_3$ perovskite structure. If the $La^{3+}$ ions (A = 1.216 Å) are partially replaced by smaller $Ce^{4+}$ ions (B ≈ 1.019 Å), t decreases. Thus, the tilt angle, $\phi$, of the oxygen octahedra also increases leading to a decrease of the hybridization among the Co3d and O2p bands and the bandwidth, W, which depends on the Co-O-Co bond angle (180° − $\phi$) through W ∝ cos$\phi$. A decreased hybridization and bandwidth also causes a decrease of the double exchange interaction which may be the reason for the strongly suppressed $T_c$.

The effective magnetic moment in the ferromagnetic state was determined from the magnetic saturation at T = 5 K and B = 5 T and amounts to $\mu_{eff}$(FM) = 0.5 $\mu_B$ per Co ion for La$_{0.7}$Ce$_{0.3}$CoO$_3$, μ$_B$ being the Bohr magneton. Since the magnetic moment of itinerant ferromagnets can be significantly reduced in the ferromagnetic state we also determined μ$_{eff}$ in the paramagnetic state from susceptibility measurements well above T$_c$, i. e., T > 50 K.[12] For La$_{0.7}$Ce$_{0.3}$CoO$_3$ we obtained μ$_{eff}$ = 3.3 μ$_B$/Co, which is much higher than the magnetic moment in the ferromagnetic state. For the hole-doped La$_{0.7}$Sr$_{0.3}$CoO$_3$, Paraskevopoulos et al. also measured a magnetic moment of μ$_{eff}$ = 3.37 μ$_B$/Co [13], indicating that the spin configurations of these electron and hole-doped cobaltites are rather similar.

From the magnetic moment in the paramagnetic state we calculated the spin value, S, assuming μ$_{eff}$ = g [S(S+1)]$^{1/2}$ μ$_B$, with a Landé factor g = 2. In Fig. 3, we show S as a function of the Ce-doping level x. We also display the S values which are expected for a LS state: t$^6_{2g}$e$^x_g$, IS state: t$^5_{2g}$e$^{1+x}_g$, and HS state: t$^{4+x}_{2g}$e$^2_g$, where S = x/2, (2+x)/2, and (4-x)/2, respectively [8], assuming a simple ionic picture. The measured S value increases with increasing doping level x, very similar to that of an IS state. The result is in agreement with a previously published theoretical work of Zhang et al., suggesting a LS-IS configuration for 0.08 < x < 0.83. In contrast to the Sr doped cobaltites where the spin state changes from LS for x < 0.25 to IS for 0.25 < x < 0.41, we did not observe any significant change in the evolution of S with x, i. e., any change in the spin state, within the doping range 0.1 ≤ x ≤ 0.4. Possibly, there is a hint discernible that the residual LS state vanishes completely above x = 0.2. Since the IS state in hole-doped cobaltites is thought to be stabilized by very strong Co3d-O2p hybridization [3], exisiting as a mixture of t$^5_{2g}$e$^1_g$ and t$^5_{2g}$e$^2_g$L$^1$, where L indicates a ligand hole, it is a rather unexpected result that the IS state in the electron-doped cobaltites seems to be so stable despite the low degree of hybridization.

In summary, we have demonstrated that it is possible to obtain electron-doped cobaltites by the growth of epitaxial thin films. The electron-doped La$_{1-x}$Ce$_x$CoO$_3$ films show ferromagnetic order within the entire doping range $0.1 \leq x \leq 0.4$ with a maximum $T_c$ of about 22 K at $x = 0.3$. With increasing Ce doping the room temperature resistivity increases indicating a decrease of hybridization among Co3d-O2p orbitals. We suggest that the decreased hybridization is caused by the decreased tolerance factor. The decreased hybridization is likely the reason for the strongly suppressed $T_c$ in comparison to the countered hole-doped cobaltites. The paramagnetic moment increases with increasing Ce-doping level x and indicates a stable IS configuration over the investigated doping range.

[11] In order to exclude possible extrinsic effects as a source for the suppressed $T_c$, we studied the influence of the oxygen partial pressure of the film deposition, the epitaxial strain and finite thickness of the films on $T_c$. We found that $T_c$ was always limited below 24 K.

[12] For a small magnetic field strength, i. e., H = 200 Oe, and $T \gg T_c$, the susceptibility, $\chi$, is given by $M/H \approx N \cdot \mu_{eff}^2/3k_BT$, where the effective magnetic moment, $\mu_{eff}$, can be determined from the slope of the magnetization M vs. 1/T. N is the number of spins/m$^3$ and $k_B$ the Boltzmann constant.

Figure Captions:

Figure 1. θ/2θ scan of $La_{0.7}Ce_{0.3}CoO_3$ using $Cu_{K\alpha}$ radiation. The substrate reflections of $LaAlO_3$ have been removed for clarity. Only 00$l$ reflections of $La_{0.7}Ce_{0.3}CoO_3$ can be observed. The inset shows an ω-scan at the 002 reflection. The full width at half maximum amounts to 0.15°.

Figure 2. Field-cooled magnetization of $La_{1-x}Ce_xCoO_3$ for x = 0.1(circles), 0.2 (down triangles), 0.3 (squares), and 0.4 (up triangles). The measurements were carried out with an external field strength of H = 200 Oe applied parallel to the film surface. The inset displays the Curie temperature, $T_c$, determined from the onset of the field-cooled magnetization as a function of the Ce doping level x.

Figure 3. The spin-value, S, calculated from $\mu_{eff}$ in the paramagnetic state, as a function of the Ce concentration x. The expected spin values for a LS state: $t^6_{2g}e^x_g$, IS state: $t^5_{2g}e^{1+x}_g$, and HS state: $t^{4+x}_{2g}e^2_g$, where S = x/2, (2+x)/2, and (4-x)/2, respectively, are displayed by solid lines.

Figures:

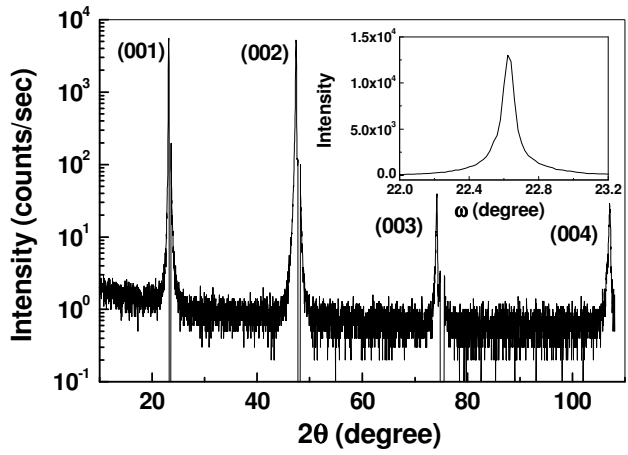

Fig. 1.

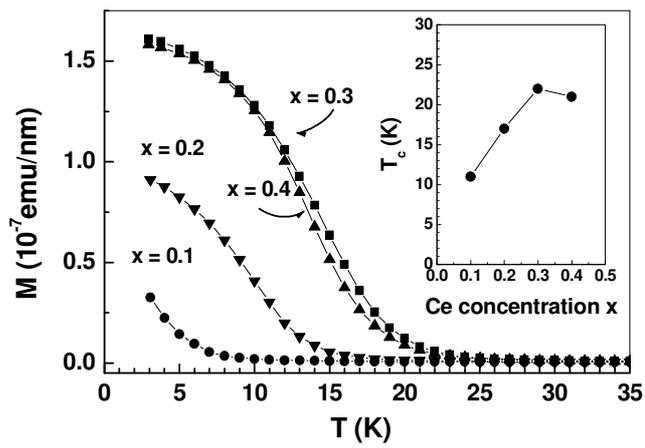

Fig. 2.

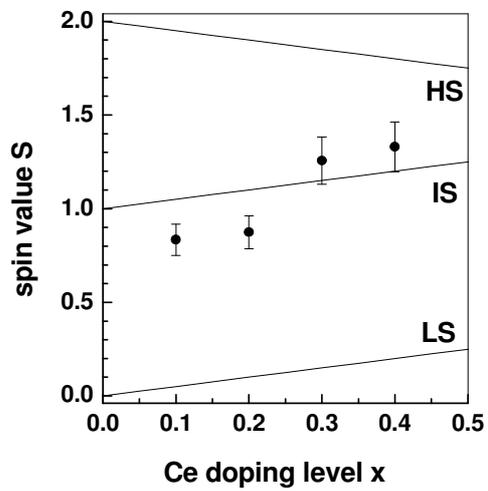

Fig. 3.

9